# On Fano effect in IR spectra of hydrogenated nanodiamonds

Andrei A. Shiryaev[a,*], Evgeni A. Ekimov[b]

[a] Frumkin Institute of physical chemistry and electrochemistry Russian Academy of Sciences, 31-4, Leninsky prospect, 119071 Moscow, Russia

[b] Vereshchagin Institute for High Pressure Physics, Russian Academy of Sciences, Troitsk, 108840 Moscow, Russia

[*] - Corresponding author: Shiryaev A.A. (shiryaev@phyche.ac.ru, a_shiryaev@mail.ru)

**Abstract.** Hydrogenated nanodiamonds may demonstrate show a “transmission window” in infra-red spectra in the vicinity of diamond Raman frequency. This phenomenon is a manifestation of resonance coupling of incident photons with continuum states (Fano resonance). However, precise mechanism of appearance of the resonance and of related conductivity – surface hydrogenation or specific type of surface reconstruction – remains debatable. We present detailed analysis of infra-red spectra of nanodiamonds of different sizes (2.6-30 nm) possessing the “transmission window” and show that the C-H stretch vibrations of adsorbed functional groups cannot be responsible for the Fano resonance. At the same time, it is suggested that a bending mode of monohydride termination on nanodiamond (111) face may couple with diamond optical phonon, explaining the Fano resonance in some cases. The relative importance of the monohydride contribution and of the graphitic islets to appearance of the “transmission window” and conductivity is likely dependent on dominating morphology and size distribution of nanodiamond grains.

## Introduction

A peculiar feature of infra-red spectra hydrogenated nanodiamonds synthesized at high static pressures and temperatures from organic compounds – the “transparency window” in the vicinity of diamond Raman frequency - was reported several years ago [1, 2]. Subsequently, it was shown that the transparency window becomes very weakly pronounced above ~200 °C upon heating both in air and in dry $N_2$ and essentially disappears by 300-350 °C [3]. The temperature effect is, however, reversible, since upon cooling to ambient conditions the absorption dip re-appears. The “transparency window”, as well as accompanying asymmetry of the Raman line, are due to Fano effect caused by resonance of the diamond Raman mode with continuum of conductive surface states. However, the main physical mechanism of its origin in nanodiamonds remains debatable. One model suggests that the conductivity is caused by specific reconstruction of broken C-C bonds and small $sp^2$-

C domains on surface of nanodiamond grains [1]. Another, more popular hypothesis, assign the effect to transfer doping mechanism, induced by adsorbed species, such as water, hydrocarbons etc. [2]. Of course, both mechanisms may operate; but the relative importance may be sample-dependent.

The Fano resonance was observed for nanodiamonds with sizes between 2.6 and 30 nm as inferred from X-ray diffraction and confirmed by Transmission Electron Microscopy [1,2]. In this letter, we reanalyze IR spectra of these nanodiamond samples with the aim to establish correlations between the amplitude, shape and fine structure of the IR dip and various peaks of C-H vibrations.

## Experimental

For detailed description of the samples' synthesis and IR spectra acquisition, see original papers [1, 2, 4]. In short, the samples were prepared from halogenated hydrocarbons at static pressure of 7.5-9 GPa at temperatures up to 1400 °C. As shown earlier, this approach allows synthesis of nanodiamonds with precisely controlled grain size by selection of a precursor and adjustment of the synthesis duration [4]. Infra-red spectra were recorded both in transflection mode with a powdered sample dispersed on Au or Al mirror and in transmission geometry with powder dispersed on a KBr pellet. For a given sample, both geometries give consistent results. In this work, transflection spectra are considered. For every sample at least three measurements in different regions were performed.

Synthesis of nanodiamonds with narrow size distribution and controlled size was described in [1]. According to X-ray diffraction patterns, the smallest nanodiamonds possess sizes below 2 nm. Strong IR absorption by C-H functional groups is observed for all studied nanodiamonds, but no Fano-related absorption dip or related distortion of the background (see below) are observed for the samples with the smallest grains. Spectra of nanodiamonds with average grain size of 2.6, 3.4, 8 and 30 nm show well-pronounced Fano-resonance and reasonable quality of spectra (Fig. 1a).

The appearance of the Fano-resonance not only induces distinct spectral features, but also distorts background. This effect is pronounced not only for nanodiamonds, but also for heavily B-doped conductive diamond films (see, e.g., Fig. 2 in [5]). Interestingly, boron-doped nanodiamonds synthesized by similar approach do not show IR absorption dip and distorted background, despite pronounced Fano-type changes in diamond Raman peak [6]. Analysis of the background behavior in terms of scattering on charge carriers ($A \sim C(\hbar\omega)^{-\gamma}$; e.g., [7]) was attempted. For the 30 nm sample $\gamma$ is ~0.3-0.4, for other samples it is close to 0. Unfortunately, in our measurements the thickness of the layer is poorly controlled, ruling out quantitative analysis. More importantly, in the case of case of nanoparticulate powder applicability of the free carriers approximation is questionable due to confinement effect. Light scattering further complicates the analysis, although its contribution to the extinction should be minor due to very small grain sizes.

Taking into account ambiguities of the eventual background slope analysis, in the current work the background was subtracted to reveal features between 1000-1500 $cm^{-1}$ and between 2700-3600 $cm^{-1}$ using polynomial fit (Fig. 1a). For simplicity, we will use the term "Fano-related" and "C-H-related" for these regions, respectively, realizing that this notation neglects contribution of C-H bending and rocking vibrations in the region 1000-1500 $cm^{-1}$. Despite some scatter in the background slopes between spectra of the same sample recorded in different locations, the "Fano"- and "C-H"-related regions are very reproducible. In some samples the C-H-related vibrations are stronger than the Fano-related ones and in some the situation is reversed. For consistency of subsequent analysis, the modulus of each background-subtracted spectra was normalized to interval [0,1]. This procedure allows for variations in relative intensities of the bands in discussed spectral intervals.

## Results

Figure 1b shows that in the small nanodiamonds with sizes ~2.6-3.4 nm the Fano-related range possess complex structure with several bands. For the larger grains (8 and 30 nm) the absorption dip is markedly asymmetric with a tail towards smaller wavenumbers; thus qualitatively resembling diamond Raman peak of nanodiamonds (e.g., [8]). For the smallest nanodiamonds the Fano region contains at least three components of roughly comparable intensity in the vicinity of the Raman frequency. These components might correspond to the Fano resonance at grains with slightly different morphology and/or size. Two components appear to be present in all samples, with somewhat different maxima positions. One of the component is at 1300-1303 $cm^{-1}$ and the second is at 1322-1329 $cm^{-1}$. The shape of Fano-related part of the sample with 8 nm grains deviates from the spectra of other samples; in it the positions of the principal Fano components are at 1285 and 1311 $cm^{-1}$. Bends of C-H vibrations may also contribute to this spectral range, but their influence appears to be of secondary importance: 1) the resonance-related "transmission window" is similar for nanodiamonds with different sizes and relative intensities of C-H-related vibrations (compare spectra of 2.6, 3.4 and 8 nm nanodiamonds, Fig. 1b); 2) eventual contribution of the bending modes would have influenced correlations between the bands in the "Fano" and "C-H" regions discussed below.

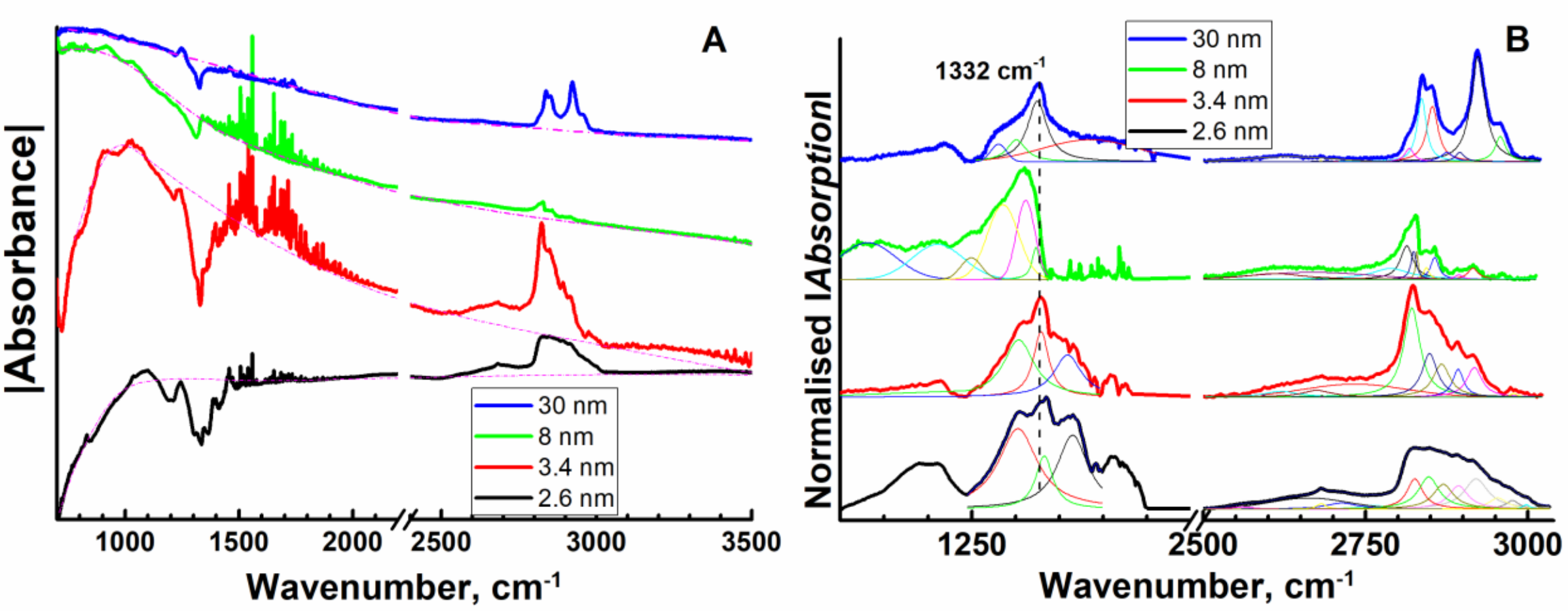

Fig. 1. Infra-red spectra of nanodiamonds with pronounced Fano-resonance features. A - representative raw spectra with example of polynomial fit of background, see text for detail. B - decomposition of spectral envelopes. Vertical axis: normalized modulus of absorption. The modulus of each spectrum is normalized to [0,1]; the curves are displaced vertically for clarity. Note break in horizontal axis.

Figure 2 shows that integrated area of the C-H related and Fano-related bands are not correlated (Pearson correlation coefficient is -0.02). Thus, analysis based on decomposition of corresponding spectral envelopes was performed. The decomposition into components was performed using Fityk software [9] assuming Lorentzian shape for all peaks except broad Gaussians; the rationale being that the later bands reflect large scatter in bond angles and lengths in poorly ordered “amorphous” matter, whereas vibrations of specific configurations of C-H moieties on various crystallographic faces of diamond grains demonstrate less scatter. The number of the components was inferred from minimal reasonable number of inflection points.

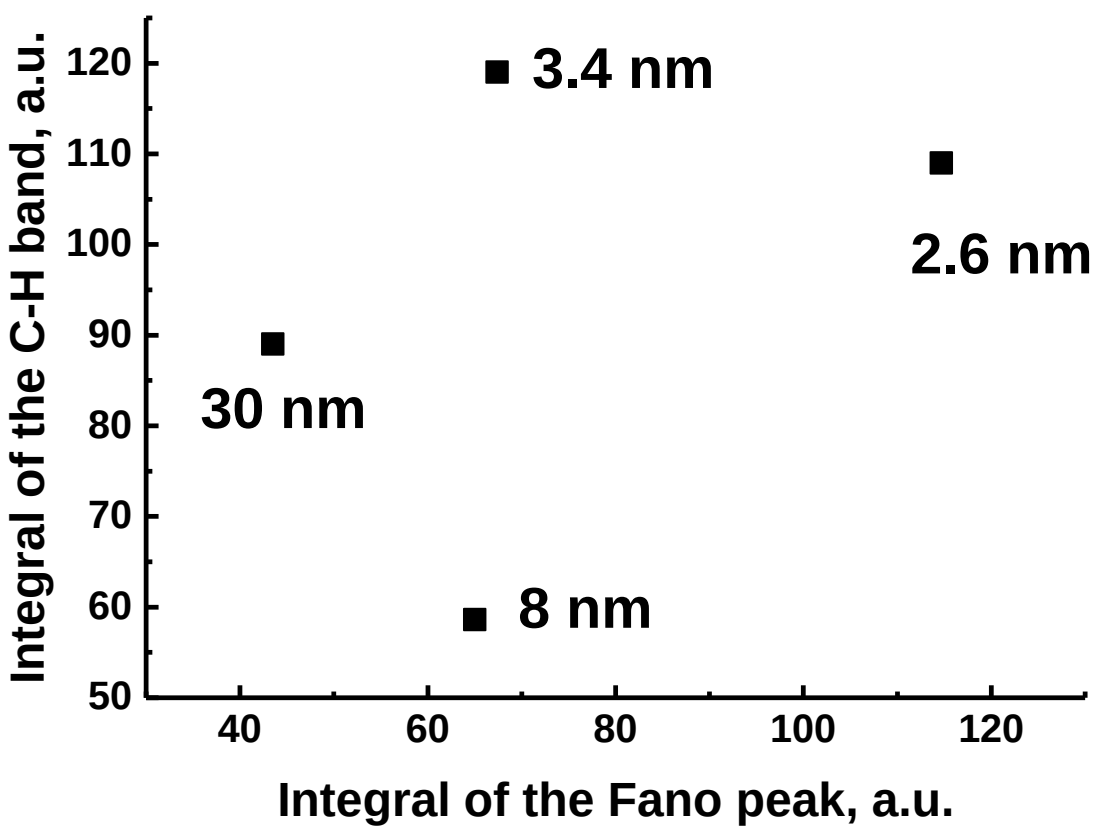

Fig. 2. Integrated area of the Fano- and C-H-related bands.

Both the Fano-related and C-H regions are rather complex and comprise numerous bands. We analyse behavior of several components obtained from the decomposition of the spectral envelope of C-H region and centered at ~2816, 2825-2837, 2847-2857, 2892-2895, 2915-2923, 2957 $cm^{-1}$. These bands are present in all spectra; position of maxima may be sample-dependent as reflected by indicated ranges. The band at 2825-2837 $cm^{-1}$ is assigned to C-H on 1×1 C(111); the 2915-2923 $cm^{-1}$ is due to C-H or $CH_2$ 2×1 C(100); 2847-2857 $cm^{-1}$ possibly corresponds to C-H on 1×1 C(110) [10-12]. The spectra reveal size-dependent differences in hydrogenation of crystal faces. Whereas the largest (30 nm) nanodiamonds show well-defined peak due to hydrogen on cube face, it is much less pronounced for the smaller grains. It is unclear whether the difference is truly size-dependent, and/or is related to growth medium and/or to presence of grains with variable morphology in the samples. We note that TEM images of the 30-nm nanodiamonds [fig. 1 in Ref. 2] show mostly octahedral grains, but the quoted images highlight grains with relatively large dimensions; the smaller grains, with larger area of (100) faces, were mentoned only briefly. In addition, it is known that the C-H bands, relatively sharp for macroscopic diamond or for nanodiamonds larger than ~100 nm, broaden considerably with decreasing nanodiamond size and reliable deconvolution becomes ambigous [13].

Examination of the IR spectra (Figs. 2 and 3) shows that the magnitude of the Fano-related peak is inversely proportional to the intensity of most C-H-related vibrations. The only possible exception is the band at ~2816 $cm^{-1}$ (see inset in Fig. 3), lacking obvious assignment. It might represent C-H stretch in aldehydes (O=C–H), but this interpretation is uncertain, since related C=O vibration is present only in spectra of the 30 nm sample (1733 $cm^{-1}$), which, in turn, shows the weakest 2816 $cm^{-1}$ band. Similar negative correlations are found for individual peaks in the Fano region (1285-1303 and 1311-1329 $cm^{-1}$). Analysis of the temperature-dependent behavior of the Fano resonance and C-H bonds vibrations for the 8 nm nanodiamonds indicates that whereas the resonance becomes weaker upon heating, the intensity of the C-H absorption, in contrast, increases [3]. The later fact is partly explained by increased IR absorption cross-section of hydrocarbons at high temperatures [14]; other factors are discussed in [3]. We note also that spectral components in the C-H-related region show only very minor variations with temperature.

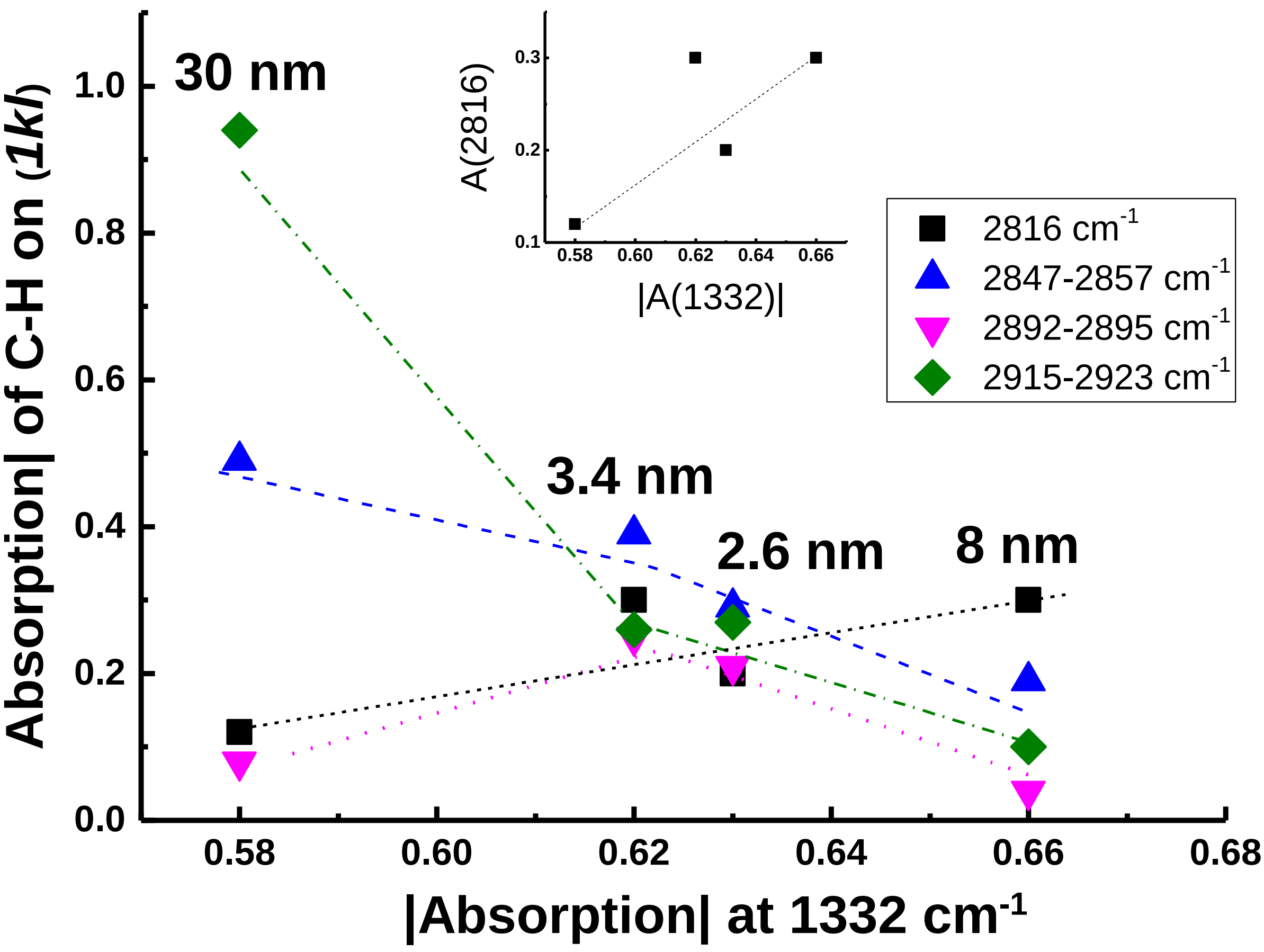

Fig. 3. Amplitude of the Fano-related infra-red "transmission window" as function of several prominent C-H bands, see text for details. The numbers indicate size of nanodiamond grains. The lines are to guide an eye. The inset shows behavior of the 2816 $cm^{-1}$ band.

## Discussion

Whereas hydrogenation definitely plays important role in surface conductivity of macro- and nanodiamond [15, 16], the present study shows that the appearance of the Fano-related transmission window, and, presumably, electrical conductivity, are unrelated to the stretch vibrations of C-H bonds on reconstructed diamond faces. This result is in line with observations for macroscopic diamond [15]. Thus, if the conductivity of hydrogenated nanodiamonds is indeed induced by adsorbed hydrocarbons, identification of the candidate species by IR appears elusive. Nevertheless, some hypotheses could be proposed. The appearance of the Fano-related IR absorption dip requires coupling between a phonon and holes. The coupling of vibrations of adsorbate species with phonons in diamond was demonstrated, for example, for deuterated hydrocarbons [8]. In the present case, a vibration close in energy to the diamond Raman frequency (1332 $cm^{-1}$) is required. Recall that IR absorption at the Raman frequency is forbidden in ideal diamond lattice, but appears upon incorporation of defects, distorting local lattice symmetry (e.g., [17]). For hydrocarbons the range 1320-1360 $cm^{-1}$ is IR-silent with few exceptions. Some $R_2$-CH bending modes; bending vibrations of OH groups in systems with strong hydrogen bonds, in-plane deformation modes of R-OH, symmetric $\nu(NO_2)$ and stretch vibrations of C-O bonds in complex ethers and lactones may be observed in this range [18,19]. Adsorption of functional groups may influence their vibration energy as manifested, for example, by clear shift of C=O stretch frequency upon adsorption on nanodiamond grains with different sizes [20]. Thus, some other functional groups with frequencies modulated by nanodiamond surface may be relevant.

A notable interesting case of a close match between a diamond phonon with an adsorbate was reported in [21], where monohydride termination of (111) diamond face was observed. The bending mode of the C-H is found at 1331 $cm^{-1}$. The corresponding stretch at 2825-2837 $cm^{-1}$ is also present in our spectra. Although its intensity does not correlate with the absolute magnitude of the Fano dip, for three samples out of four (the 8 nm is an exception), almost linear relationship is observed between this band and a component peaking at 1322-1329 $cm^{-1}$ (Fig. 4). We recall here that the Fano-related features form a "transmission window", whereas the C-H-related bands absorb. Thus, the positive correlation is not just an indication of the monohydride presence, but, rather, strongly suggest that the relevant C-H bend vibration is indeed related to the appearance of the IR absorption dip. The temperature-induced shift of the bending mode frequency may lead to detuning of the hole-phonon resonance and gradual disappearance of the absorption dip and its recovery upon cooling.

Marked deviation of the 8 nm sample from the trend mentioned above is yet unexplained. However, the spectral envelope of the Fano region for this sample differs from the other studied nanodiamonds. It might be a manifestation of sample-dependent scatter in shapes and/or size distribution of the grains, but precise reasons remain unkown.

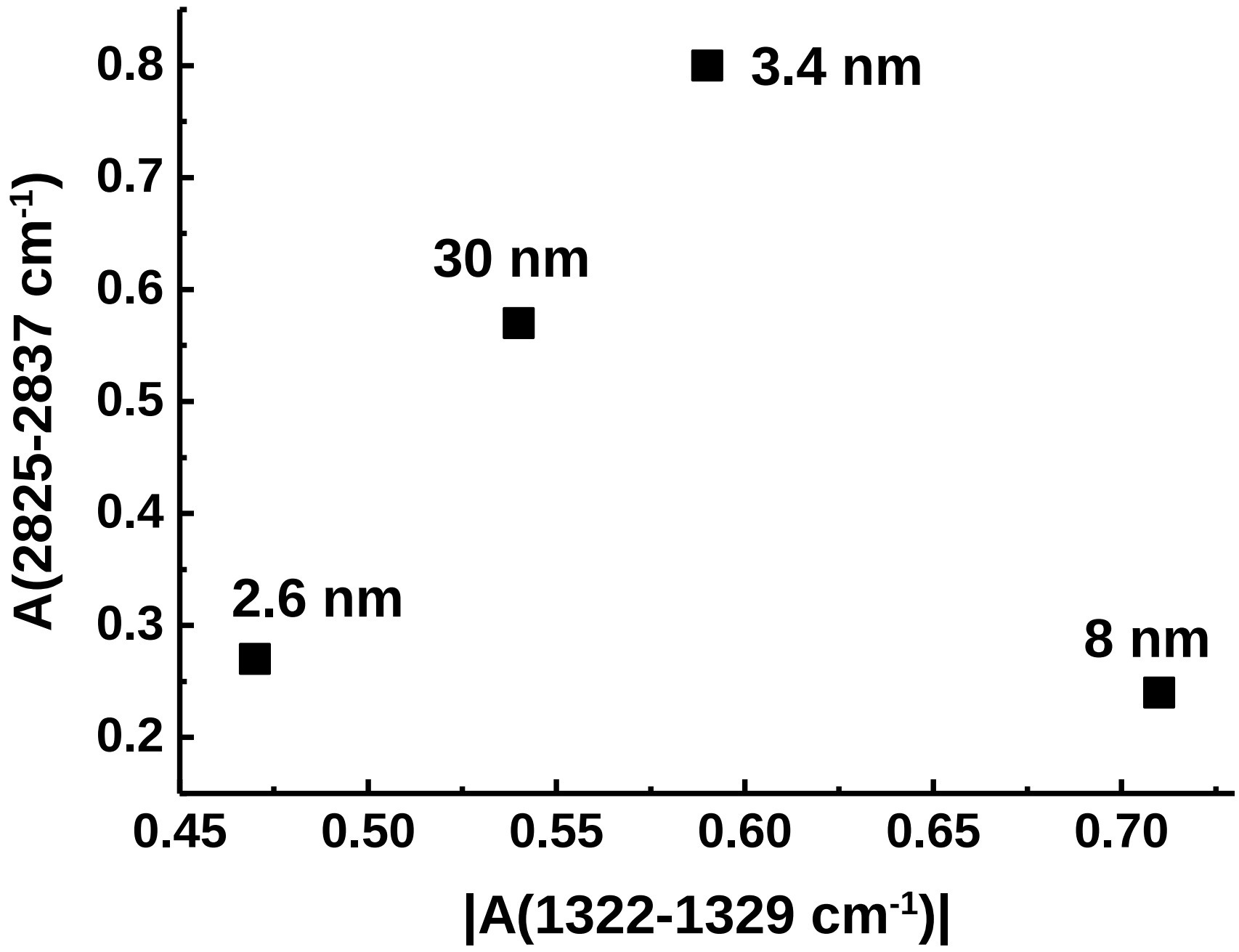

Fig. 4. Amplitude of the 2825-2837 $cm^{-1}$ band as a function of the modulus of absorption of a component peaking at 1322-1329 $cm^{-1}$. The numbers indicate size of nanodiamond grains.

IR spectra of the studied nanodiamonds contain a small, but distinct absorption peak at 2682 cm$^{-1}$ (Fig. 5). The intensity of this peak sharply decreases with increasing size of nanodiamonds, being the strongest in the 2.4 nm specimen. Unique assignment of this feature is not yet possible, but several possibilities could be considered. It may correspond to C-H stretch modes in CHO groups [19] or to C-H stretch on graphite surface [22]. The later hypothesis is reconcilable with higher fraction of graphitic carbon for smaller nanodiamonds. This, in turn, is consistent with appearance of the required electrical conductivity due to graphitic islets related to peculiarities of surface reconstruction [1].

Interestingly, position of this feature is only marginally higher than doubled Raman frequency (2×1332 cm$^{-1}$ = 2664 cm$^{-1}$) of diamond. At the same time, phonon density of states of nanodiamonds extends slightly higher than for a macrocrystal [23]. The present work indicates that in nanodiamonds optical phonons may couple (directly or indirectly) with incident IR radiation; thus, interaction with two phonons is not excluded.

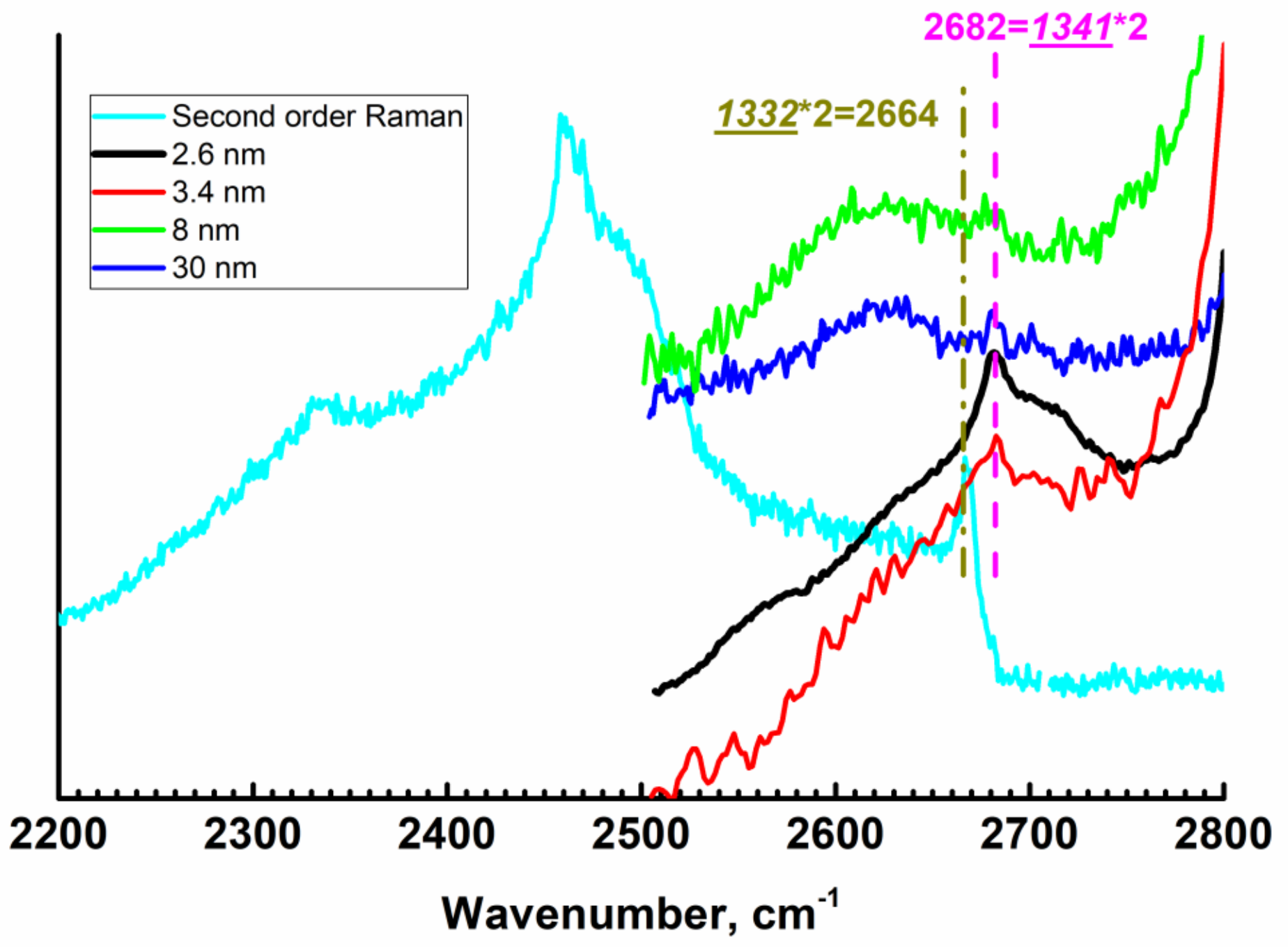

Fig. 5. Zoomed region of IR spectra showing possible correspondence between observed features and phonon density of states. The presented second order Raman spectrum was recorded for single crystal diamond.

**Conclusions.**

Analysis of infra-red spectra of hydrogenated nanodiamonds showing "transmission window" at diamond Raman frequency is presented. The absorption dip is related to Fano-type resonance between a phonon and a hole. Two principal phenomena may be responsible for the conductivity: 1) transfer doping induced by adsorbed hydrocarbons, and 2) formation of graphitic islets and specific type of surface reconstruction. Here we show that stretch vibrations of functional C-H groups on nanodiamond faces cannot be responsible for the resonance. However, it is plausible, monohydride bending mode on the (111) nanodiamond faces may couple with the diamond phonon and be instrumental for the Fano resonance. At the same time, small absorption features tentatively assigned to C-H modes on graphitic domains are observed and are especially pronounced for the smallest studied nanodiamonds. Thus, both mechanisms – transfer doping assisted by monohydride (111) termination and presence of graphitic islets on reconstructed surfaces – may be responsible for the appearance of surface conductivity of nanodiamonds; the relative contribution of each mechanism appears to be dependent on peculiarities of nanodiamond grain shapes and size distribution in a given sample.

**Acknowledgements**

We appreciate useful comments of an anonymous reviewer. The work was funded by the Ministry of Science and Higher Education of the Russian Federation. IR measurements were performed using equipment of CKP FMI IPCE RAS.

**Author contributions: CRediT**

AAS – conceptualization, methodology, investigation, supervision, writing; EAE – sample preparation, writing. Both authors have read the manuscript and agree with its content.

**References**

[1] E. Ekimov, A.A. Shiryaev, Y. Grigoriev, A. Averin, E. Shagieva, S. Stehlik and M. Kondrin. Size-Dependent Thermal Stability and Optical Properties of Ultra-Small Nanodiamonds Synthesized under High Pressure, Nanomaterials (2022), **12**(3), 351, https://www.mdpi.com/2079-4991/12/3/351/pdf

[2] O.S. Kudryavtsev, R.H. Bagramov, A.M. Satanin, A.A. Shiryaev, O.I. Lebedev, A.M. Romshin, D.G. Pasternak, A.V. Nikolaev, V.P. Filonenko, and I.I. Vlasov. Fano-type Effect in Hydrogen-Terminated Pure Nanodiamond, Nano Letters (2022), 22, 2589−2594, https://doi.org/10.1021/acs.nanolett.1c04887

[3] A. A. Shiryaev, E. A. Ekimov, V. Yu. Prokof'ev, and M. V. Kondrin. Temperature Dependence of the Fano Resonance in Nanodiamonds Synthesized at High Static Pressures, JETP Letters (2022), 115(11), 651–656

[4] E. Ekimov, S. Lyapin, Y. Grigoriev, I. Zibrov, K. Kondrina. Size-controllable synthesis of ultrasmall diamonds from halogenated adamantanes at high static pressure. Carbon (2019), 150, 436–438.

[5] J. W. Ager III, W. Walukiewicz, M. McCluskey, M. A. Plano, M. I. Landstrass. Fano interference of the Raman phonon in heavily boron-doped diamond films grown by chemical vapor deposition. Applied Physics Letters (1995), 66, 616; doi: 10.1063/1.114031

[6] E.A. Ekimov, A.A. Shiryaev, T.B. Shatalova, K.I. Maslakov, S.G. Lyapin, K.M. Kondrina, Y.V. Grigoriev, S. Stehlik, M.V. Kondrin. Thermal stability and oxidation resistance of single-digit boron-doped nanodiamonds, Materials Research Bulletin (2025), 192, 113604

[7] H. Y. Fan. Effects of Free Carriers on the Optical Properties. Semiconductors and semimetals (1967), 3, 405-419.

[8] S. Osswald, V. N. Mochalin, M. Havel, G. Yushin, Y. Gogotsi. Phonon confinement effects in the Raman spectrum of nanodiamond. Physical review B (2009), 80, 075419.

[9] M. Wojdyr. Fityk: a general-purpose peak fitting program. J. Appl. Cryst. (2010) 43, 1126-1128

[10] H.-C. Chang, J.-C. Lin; J.-Y. Wu, K.-H. Chen. Infrared Spectroscopy and Vibrational Relaxation of CH, and CD, Stretches on Synthetic Diamond Nanocrystal Surfaces, J. Phys. Chem. (1995), 99, 11081-11088

[11] C.-L. Cheng, J.-C. Lin, and H.-C. Chang. The absolute absorption strength and vibrational coupling of CH stretching on diamond C(111), The Journal of Chemical Physics (1997), 106, 7411; doi: 10.1063/1.473701

[12] J.-C. Lin, K.-H. Chen, H.-C. Chang, C.-S. Tsai, C.-E. Lin, and J.-K. Wang. The vibrational dephasing and relaxation of CH and CD stretches on diamond surfaces: An anomaly, J. Chem. Phys. (1996), 105, 3975.

[13] S.-Y. Sheu, I.-P. Lee, Y. T. Lee, and H.-C. Chang. Laboratory investigation of hydrogenated diamond surfaces: implications for the formation and size of interstellar nanodiamonds, The Astrophysical Journal (2002), 581, L55–L58.

[14] A. E. Klingbeil, J. B. Jeffries, R. K. Hanson. Temperature-dependent mid-IR absorption spectra of gaseous hydrocarbons. Journal of Quantitative Spectroscopy & Radiative Transfer (2007), 107 407–420,

[15] F. Maier, M. Riedel, B. Mantel, J. Ristein, L. Ley. Origin of surface conductivity in diamond. Phys. Rev. Lett. (2000), 85, 3472−3475.

[16] A. Bolker, C. Saguy, R. Kalish. Transfer doping of single isolated nanodiamonds, studied by scanning probe microscopy techniques. Nanotechnology (2014), 25, 385702.

[17] J. L. Birman. Theory of Crystal Space Groups and Infra-Red and Raman Lattice Processes of Insulating Crystals. In: Theory of Crystal Space Groups and Lattice Dynamics, 1–521. Berlin, Heidelberg: Springer. (1974) https://doi.org/10.1007/978-3-642-69707-4_1

[18] J.C.D. Brand, G. Eglington. Applications of spectroscopy to organic chemistry. Oldbourne press, London (1965)

[19] D.W. Brown, A.J. Floyd, M. Sainsbury. Organic spectroscopy. Wiley. (1988)

[20] J.-S. Tu, E. Perevedentseva, P.-H. Chung, and C.-L. Cheng. Size-dependent surface CO stretching frequency investigations on nanodiamond particles, J. Chem. Phys. (2006), 125, 174713; doi: 10.1063/1.2370880

[21] R. P. Chin, J. Y. Huang, and Y. R. Shen, T. J. Chuang, H. Seki. Interaction of atomic hydrogen with the diamond C(111)surface studied by infrared-visible sum-frequency-generation spectroscopy. Physical Review B (1995), 52(8), 5985-5995

[22] A. Allouche, Y. Ferro, T. Angot, C. Thomas, J.-M. Layet. Hydrogen adsorption on graphite (0001) surface: A combined spectroscopy–density-functional-theory study. J. Chem. Phys. (2005), 123, 124701; doi: 10.1063/1.2043008

[23] A.A. Shiryaev, V.B. Polyakov, S. Rols, A. Rivera, O.A. Shenderova. Inelastic neutron scattering: A novel approach towards determination of equilibrium isotopic fractionation

factors. Size effects on heat capacity and beta-factor of diamond, Physical Chemistry Chemical Physics (2020), **22**, 13261 – 13270, DOI: 10.1039/D0CP02032J